\documentclass[11pt]{elsart}
\usepackage[dvips]{graphicx}

\begin{document}
\begin{frontmatter}

\title{Boolean Game on Scale-free Networks}

\author{Jing Ma$^{a,b}$ }
\author{Pei-Ling Zhou$^{a}$ }
\author{Tao Zhou$^{a,c}$}
\footnote{corresponding author: zhutou@ustc.edu}
\author{Wen-Jie Bai$^{d}$ }
\author{Shi-Min Cai$^{a}$ }

\address
{$^{a}$ Department of Electronic Science and Technology,
University of Science and Technology of China,  Hefei Anhui,
230026, PR China}

\address
{$^{b}$ Department of Mathematics, University of Science and
Technology of China,  Hefei Anhui, 230026, PR China}

\address
{$^{c}$ Department of Modern Physics and Nonlinear Science Center,
University of Science and Technology of China,  Hefei Anhui,
230026, PR China }

\address
{$^{d}$ Department of Chemistry, University of Science and
Technology of China,  Hefei Anhui, 230026, PR China}

\begin{abstract}
Inspired by the local minority game, we propose a network Boolean
game and investigate its dynamical properties on scale-free
networks. The system can self-organize to a stable state with
better performance than random choice game, although only the
local information is available to the agent. By introducing the
heterogeneity of local interactions, we find the system has the
best performance when each agent's interaction frequency is linear
correlated with its information capacity. Generally, the agents
with more information gain more than those with less information,
while in the optimal case, each agent almost has the same average
profit. In addition, we investigate the role of irrational factor
and find an interesting symmetrical behavior.

\begin{keyword}
Boolean Game, Local Minority Game, Scale-Free Networks,
Self-Organization \PACS 02.50.Le, 05.65.+b, 87.23.Ge, 89.75.Fb
\end{keyword}
\end{abstract}

\date{}
\end{frontmatter}

%%%%%%%%%%%%%%%%%%%%%%%%%%%%%%%%%%%%%%%%%%%%%%%%%%%%%%%%%%%%%%%%%%%%
\section{Introduction}
In recent years, the phenomena of collective behavior related to
populations of interacting individuals attract increasing
attentions in the studies on scientific world, especially in the
economical and biological systems
\cite{Cambridge1,Cambridge2,begin}. To describe and explain the
self-organized phenomenon, many models are established. Inspired
by Arthur's Farol Bar Problem \cite{Originin}, Challet and Zhang
proposed the so-called minority game (MG)
\cite{Original1,Original2}, which is a simple but rich model
showing how selfish agents fight for common resources in the
absence of direct communication.

In the standard minority game, a group of $N$ (odd) agents has to
choose between two opposing actions, which are labelled by $+1$
and $-1$ , respectively. In the real systems of stock market,
these options mean to buy stock or to sell. Each agent is assigned
a set of $s$ strategies and informed the updated global outcomes
for the past $M$ time steps. At each time step, they use the most
working strategies to make decisions, and those who end up in the
minority side (the side chosen by fewer agents) win and get a
point. Though simple, MG displays the self-organized
global-cooperative behaviors which are ubiquitously observed in
many social and economic systems
\cite{StudiedMG1,StudiedMG2,StudiedMG3,StudiedMG4,StudiedMG5,StudiedMG6,StudiedMG7,StudiedMG8,StudiedMG9,StudiedMG10,Quan2002}.
Furthermore, it can explain a large amount of empirical data and
might contribute to the understanding of many-body ecosystems
\cite{ModifiedMG2,ModifiedMG3,ModifiedMG4}.

In the real world, an individual ia able to get information from
his/her acquaintances, and try to perform optimally in his/her
immediate surroundings. In order to add this spatial effect to the
basic MG, recently, some authors introduced the so-called local
minority game (LMG), where agent could make a wiser decision
relying on the local information \cite{rnet
andLMG1,LMG3,LMG4,LMG5,LMG6,LMG7,LMG9}. It is shown that the
system could benefit from the spatial arrangement, and achieves
self-organization which is similar to the basic MG.

Denote each agent by a node, and generate a link between each pair
of agents having direct interaction, then the mutual influence can
vividly be described by means of the information networks.
Accordingly, node degree $k$ is proportional to the quantity of
information available to the corresponding agent. Most LMG models
are based on either the regular networks, or the random ones.
Nevertheless, both of them have a characterized degree, the mean
degree $\langle k \rangle$, which means each agent is almost in
the same possession of information. However, previous studies
reveal that the real-life information networks are highly
heterogeneous with approximately power-law distributions
\cite{BAproperty1,BAproperty2,BAproperty3}. Thus the above
assumption is quite improper for the reality. In common sense,
those who process huge sources of information always play active
and important roles. Therefore, in this paper, we will study the
case on the base of scale-free networks.

Another interesting issue is the herd behaviors that have been
extensively studied in \textbf{Behavioral Finance} and is usually
considered as a crucial origin of complexity that enhances the
fluctuation and reduces the system profit
\cite{herdbehavior1,herdbehavior2,herdbehavior3,herdbehavior4,herdbehavior5,herdbehavior6}.
Here we argue that, to measure the potential occurrence of herd
behavior, it is more proper to look at how much an agent's actions
are determined by others (i.e. the local interaction strength of
him) rather than how much he wants to be the majority. It is
because that in many real-life situations, no matter how much the
agents want to be the minority, the herd behavior still occurs. To
reveal the underlying mechanism of the herd behavior, three
questions are concerned in this paper:

a) Whether agents have different responses under the same
interaction strength?

b) What are the varying trends of individual profit as the
increase of interaction strength?

c) What are the effects of heterogenous distribution of individual
herd strength on system profit?

Furthermore, a fundamental problem in complexity science is how
large systems with only local information available to the agents
may become complex through a self-organized dynamical process
\cite{RNetwork,RNetworkhou}. In this paper, we will also discuss
this issue based on the present model by detecting the
profit-degree correlations.

\section{Model}
In the present model, each agent chooses between two opposing
actions at each time step, simplified as +1 and -1. And the agents
in the minority are rewarded, thus the system profit equals to the
number of winners \cite{Original2,StudiedMG3}. At each time step,
each agent $x$ will, at probability $p_x$, make a call to one
randomly selected neighbor to ask about this neighbor's last
action, and then decide to choose the opposing one; or at
probability $1-p_x$, agent $x$ simply inherits his previous
action. Accordingly, in the former case, agent $x$ will choose +1
at a probability $\frac{s_{-1}^{x}}{s_{-1}^{x}+s_{+1}^{x}}$, or
choose -1 at a probability
$\frac{s_{+1}^{x}}{s_{-1}^{x}+s_{+1}^{x}}$, where $s_{+1}^{x}$ and
$s_{-1}^{x}$ denote the number of $x$'s neighbors choosing +1 and
-1 in the last time step, respectively. It is worthwhile to
emphasize that the agents do not know who are the winners in the
previous steps since the global information is not available. This
is one of the main distinctions from the previously studied LMG
models.

Take the irrational factor into account \cite{herdbehavior4}, each
agent may, at a mutation probability $m$, choose an opposite
action. The mutation probability adds several impulsive and
unstable ingredients to our model. Just as the saying goes,
`nothing is fixed in the stone', actually, people can not consider
every potential possibility that would come out when making a
decision. So, it is the case that we are making the mind at this
moment and changing our mind at the next. To this extent, the
introduction of the mutation parameter enriches the model.

\begin{figure}
\begin{center}
\scalebox{1.2}[1.2]{\includegraphics{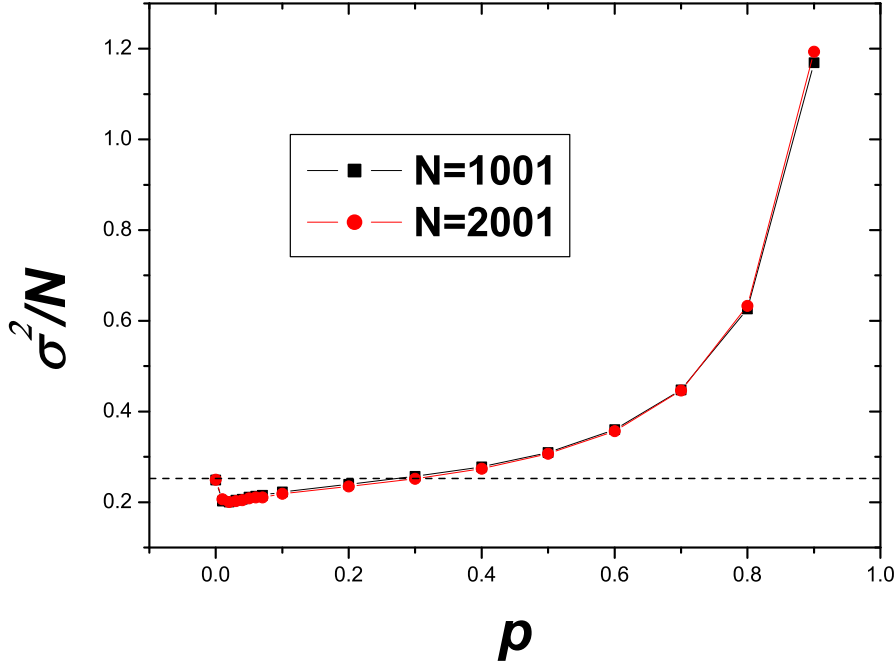}} \caption{(Color
online) The normalized variance as a function of the average
interaction strength $p$ on the BA network of size 1001 and size
2001. The dashed line represents the system profit of random game.
The system performs better than the random game when
$p\in(0,0.28)$. Besides, the two curves are almost the same,
indicating the present $\frac{\sigma^2}{N} - p$ relationship does
not depend on the system size $N$.}
\end{center}
\end{figure}

\begin{figure}
\begin{center}
\scalebox{1.1}[1.1]{\includegraphics{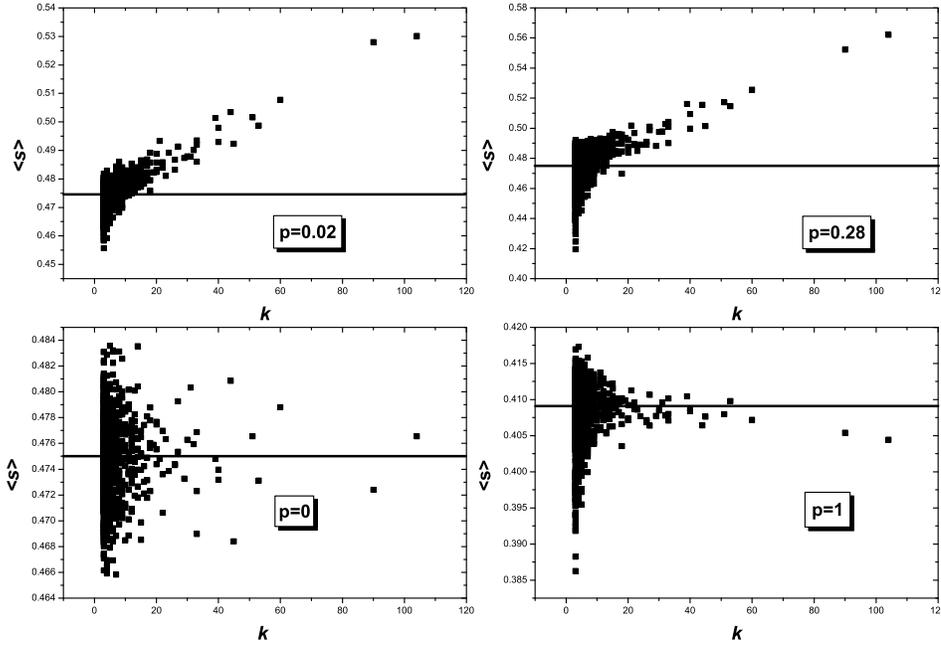}}\caption{The agent's
winning rate versus degree. Each data point denotes one agent and
the solid line represents the average winning rate of all the
agents. The cases $p=0$ and $p=1.0$ correspond to the completely
independent and dependent cases, respectively; $p=0.02$ is the
point where the system performs best, and $p=0.28$ is another
point where the system profit is equal to the random game.}
\end{center}
\end{figure}

Considering the potential relationship between individual's
information capacity and herd strength, we assume $p_x\sim
k_x^\beta$, where $k_x$ is $x$'s degree, and $\beta$ is a free
parameter. Denote $p=\frac{1}{N}\sum_xp_x$ the average herd
strength, then one has
\begin{equation}
p_x=Np\frac{k_x^\beta}{\sum_yk_y^\beta},
\end{equation}
where the subscript $y$ goes over all the nodes.

There are three cases for interaction strength distributions.

a)  $\beta=0$, each agent of the network shares the same
interaction strength $p$.

b)  $\beta>0$, heterogeneity occurs: The greater the degree, the
stronger the interaction strength, that is, hub nodes depend on
the local information while small nodes exhibit relatively
independent decision making.

c)  $\beta<0$, the heterogeneity occurs in the opposite situation:
The smaller the degree, the stronger the interaction strength,
that is, hub nodes exhibit independence while small nodes depend
on the local information.

The special case with $m=0.01$ and $\beta=0$ has been previously
studied to show the effect of degree heterogeneity on the
dynamical behaviors \cite{Zhou2005}.

\section{Simulation and Analysis}

\subsection{Self-organized phenomenon}
In this paper, all the simulation results are averaged over 100
independent realizations, and for each realization, the time
length is $T=10000$ unless some special statement is addressed.
The Barab\'asi-Albert (BA) networks with minimal degree $m=3$ are
used \cite{BA1,BA2}. Initially, each node randomly choose $+1$ or
$-1$. In this subsection, we concentrate on the case $\beta=0$ and
$m=0.01$.

\begin{figure}
\begin{center}
\scalebox{1.2}[1.2]{\includegraphics{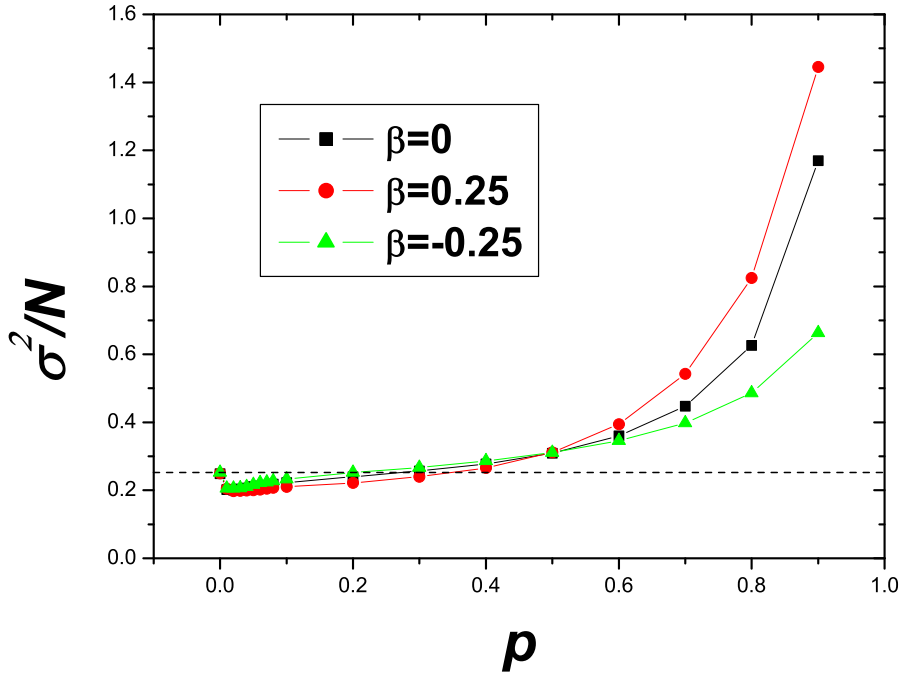}} \caption{(Color
online) The normalized variance as a function of the average
interaction strength $p$. The dashed line represents the system
profit of random game. The system performs better than random game
when $p\in(0,0.2)$, $p\in(0,0.28)$ and $p\in(0,0.34)$ for
$\beta=-0.25$, $\beta=0$ and $\beta=0.25$, respectively. }
\end{center}
\end{figure}

The performance of the system can be measured by the variance
$\sigma^{2}=(1/T)\Sigma_{t=1}^{T}(A_t-N/2)^{2}$, where $A_t$ is
the number of agents who choose +1 at time step t, and $N$ denotes
the network size \cite{Original2,StudiedMG3,Quan2002}. Clearly,
smaller $\sigma^{2}$ corresponds to more system profit and for the
completely random choice game (random game for short),
$\sigma^{2}=0.25N$. Fig. 1 shows the normalized variance
$\frac{\sigma^2}{N}$ as a function of the average interaction
strength $p$ (Since $\beta=0$, all the nodes have the same
interaction strength $p$). Unexpectedly, although the global
information is unavailable, the system is able to perform better
than random game in the interval $p \in (0,0.28)$. This is a
strong evidence for the existence of self-organized process. In
addition, we attain that the network size effect is very slight
for sufficient $N$ ($N \sim 10^3$), thus hereinafter, only the
case $N=1001$ is investigated.

In Fig. 2, we report the agent's winning rate versus degree, where
the winning rate is denoted by the average score $\langle s
\rangle$ during one time step. Obviously, unless the two extreme
points $p=0$ and $p=1$, there is a positive correlation between
the agent's profit and degree, which means the agents of larger
degree will perform better than those of less degree. If the
agents choosing $+1$ and $-1$ are equally mixed up in the network,
there is no correlation between profit and degree \cite{Zhou2005}.
Therefore, this positive correlation provides another evidence of
the existence of a self-organized process.

\subsection{Effect of interaction strength heterogeneity on system
profit}

In this subsection, we investigate how $\beta$ affects the system
profit with mutation probability $m=0.01$ fixed.

In Fig. 3, it is observed that in all the three cases
$\beta=-0.25, \beta=0,$ and $\beta=0.25$, the system performs more
efficiently than the random choice game when $p$ is at a certain
interval. More interesting, when the interaction strength $p$ is
small ($p<0.5$), the system with positive $\beta$ ($\beta=0.25$)
performs best, while for large $p$ ($p>0.5$), the system with
negative $\beta$ ($\beta=-0.25$) performs best. However, this
phenomenon does not hold if $|\beta|$ is too large ($\beta\cong
1$).

We have checked that for all the cases with $\beta \in (-1,1)$,
all the systems achieve their own optimal state at the interval
$p\in[0.02,0.1]$. When given $p$, it is natural to question
whether there exists an optimal $\beta$, in which the system
performs best. We report the normalized variance as a function of
$\beta$ for different $p$ in figure 4. Remarkably, all the optimal
states are achieved around $\beta=1$. Besides, it is worthwhile to
attach significant importance to the case when $\beta=1$ and
$p=0.08$, when we have the most profitable system.

\begin{figure}
\begin{center}
\scalebox{1.2}[1.2]{\includegraphics{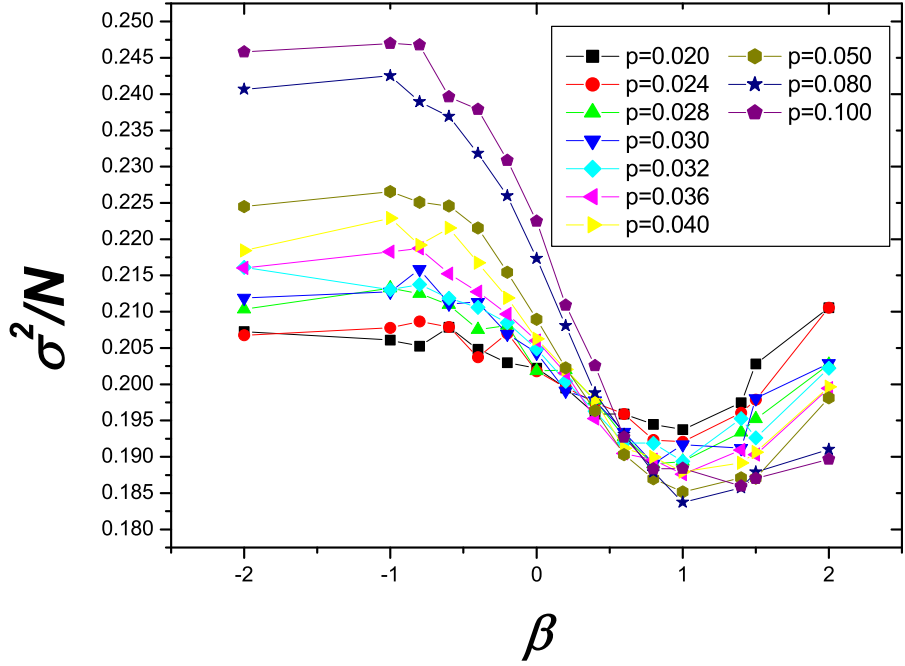}}   \caption{(Color
online)  the normalized variance as a function of the interaction
power $\beta$ for several selected $p$ indicated in the inset.
Evidently, the system tends to perform optimally when the
interaction power $\beta$ is
 around 1. }
\end{center}
\end{figure}

\begin{figure}
\begin{center}
\scalebox{1.2}[1.2]{\includegraphics{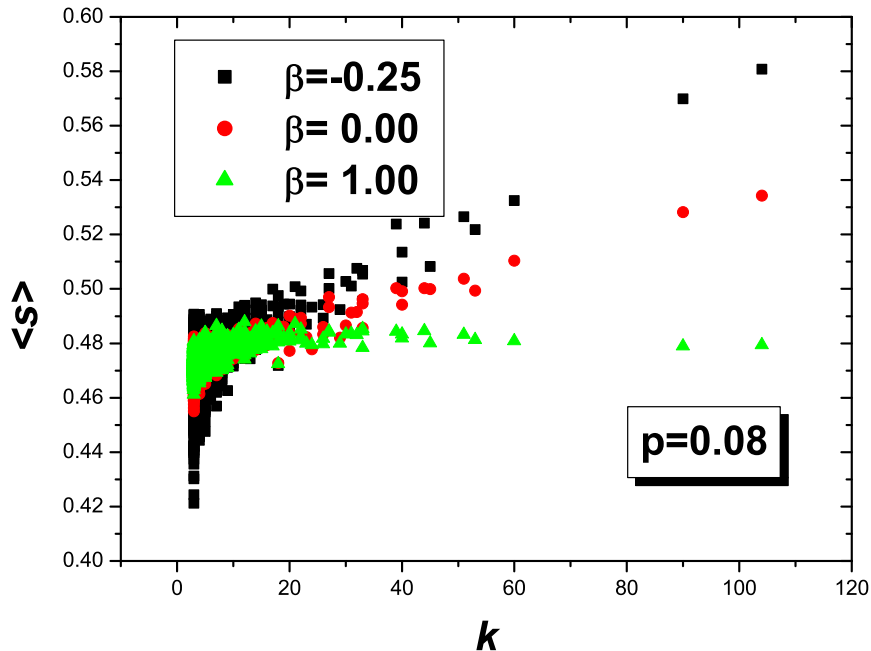}} \caption{(Color
online) The agent's winning rate versus degree, where each data
point corresponds to one agent. The black squares, red circles and
green triangles denote the case $\beta=-0.25$, $\beta=0$ and
$\beta=1$, respectively.}
\end{center}
\end{figure}

\begin{figure}
\begin{center}
\scalebox{1.2}[1.2]{\includegraphics{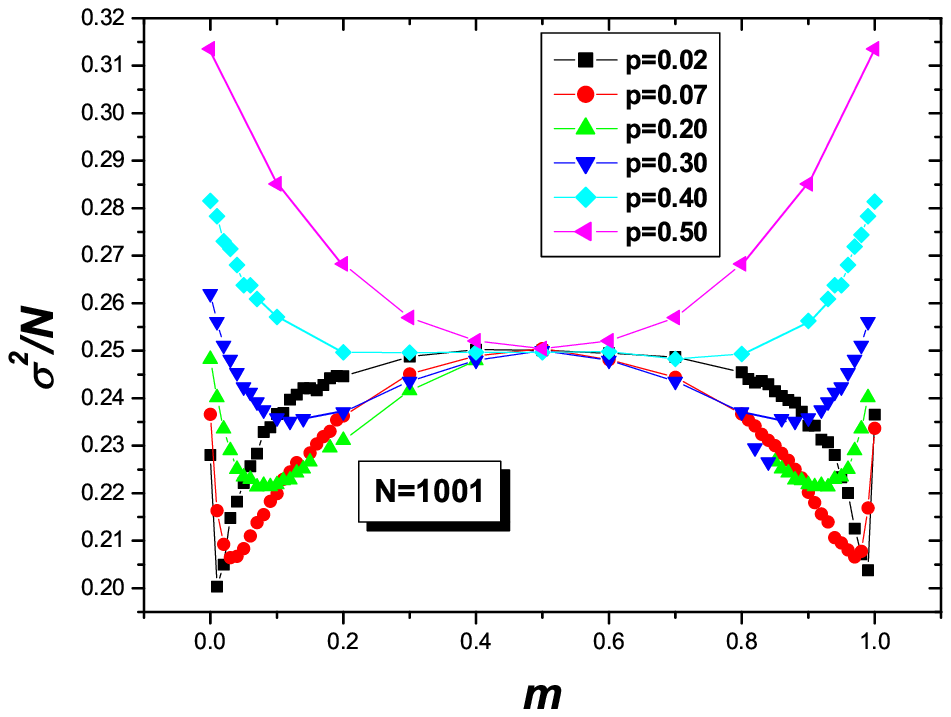}}\caption{(Color
online) The normalized variance as a function of the mutation
probability $m$ for several selected $p$. The parameter $\beta=0$
is fixed. Obviously, the curves exhibit symmetry with a dividing
point $m=0.5$. When $m=0.5$, every system has the same profit as
the random game though different $p$. The shapes of these curves
vary as $p$ increases: For small $p$, the curves have the peculiar
``two-winged" shape, while for large $p$, the curves become
U-shaped. The cases for $p>0.5$ have very large $\sigma^2/N$ when
$m$ is close to 0 or 1, thus not shown here.}
\end{center}
\end{figure}

The interaction strength heterogeneity can have some positive
effects on system as a whole, including better use of the
information and more profit. We wonder which group of people
profit after all? Specifically, figure 5 shows the agent's winning
rate versus degree for $\beta=-0.25$, 0 and 1, where $p=0.08$ is
fixed. Unexpectedly, in the most optimal system (i.e. $\beta=1$
and $p=0.08$), the profit-degree correlation vanishes. So one may
draw an interesting conclusion that the great disparity between
poor and rich population is not the necessary condition for an
efficient social. However, we can not give an analytical solution
about this phenomenon, and the corresponding conclusion may be
only valid for this special model.

\subsection{Role of irrational factor}
Figure 6 reports the normalized variance as a function of the
mutation probability $m$. Interestingly, the curves display
symmetry with a dividing point $m=0.5$ at which each system has
the same profit as the random game. For arbitrary agent, denote
$s_a$ the number of this agent's neighbors choosing $a$ in the
present time step, and $p_{(a,b)}$ the probability he/she will
choose $b$ in the next time step under the condition that he/she
chooses $a$ in the present time step. Clearly, one has
\begin{equation}
p_{(+1,+1)}=\frac{(1-m)p\cdot s_{-1}}{s_{-1}+s_{+1}}+\frac{m\cdot
p\cdot s_{+1}}{s_{+1}+s_{-1}}+(1-m)(1-p)\\
\end{equation}
\begin{equation}
p_{(+1,-1)}=\frac{(1-m)p\cdot s_{+1}}{s_{-1}+s_{+1}}+\frac{m\cdot
p\cdot s_{-1}}{s_{+1}+s_{-1}}+m(1-p)
\end{equation}
\begin{equation}
p_{(-1,+1)}=\frac{(1-m)p\cdot s_{-1}}{s_{-1}+s_{+1}}+\frac{m\cdot
p\cdot s_{+1}}{s_{+1}+s_{-1}}+m(1-p)
\end{equation}
\begin{equation}
p_{(-1,-1)}=\frac{(1-m)p\cdot s_{+1}}{s_{-1}+s_{+1}}+\frac{m\cdot
p\cdot s_{-1}}{s_{+1}+s_{-1}}+(1-m)(1-p)
\end{equation}

If $m=0.5$, $p_{(+1,+1)}=p_{(+1,-1)}=p_{(-1,+1)}=p_{(-1,-1)}=0.5$
(the same as that of the random game), thus $\sigma^{2}/N=0.25$
(independent of $p$). Additionally, replace $m$ by $1-m$, one will
immediately find the symmetry.

\section{Conclusion}
In summary, inspired by the local minority game, we propose a
network Boolean game. The simulation results upon the scale-free
network are shown. The system can self-organize to a stable state
with a better performance than the random choice game, although
only the local information is available to the agent. This is a
reasonable evidence of the existence of a self-organized process.
We find remarkable differences between the case with local
interaction strengths identical for all agents ($\beta=0$), and
that with local interaction strengths unequally distributed to the
agents. The interval of $p$, within which the system can perform
better than the random game, is obviously extended in the case
when $\beta>0$. In addition, the system reaches the best
performance when each agent's interaction frequency is linear
correlated with its information capacity. Generally, the agents
with more information gain more, however, in the optimal case,
each agent has almost the same average profit. Within the frame of
this model, the great disparity between poor and rich population
is not the necessary condition for an efficient social. The effect
of irrational factor on the dynamics of this model is also
investigated, and an interesting symmetrical behavior is found.

Although is rough, the model offers a simple and intuitive
paradigm of many-body systems that can self-organize even when
only local information is available. Since the self-organized
process is considered as one of the key ingredients of the origins
of complexity, hopefully, the model as well as its perspectives
and conclusions might contribute to the achievement of the
underlying mechanism of the complex systems. Furthermore, using
the method proposed by Challet, Marsili, and Zhang
\cite{CMZ1,CMZ2,CMZ3}, the market price can be introduced to the
present model by define the price return proportional to the
difference between the numbers of agents choosing $+1$ and $-1$.
In this sense, one is able to check if this model can display
\emph{stylized facts} in accordance with the real markets.

Finally, we would like to point out that to set some kinds of
action strength correlated with the degree of corresponding node
in a power-law form (e.g. $p_x\sim k_x^\beta$) to better the
system performance is not only available in this particular issue,
but also a widely used approach for many dynamics upon scale-free
networks, such as to fasten searching engine \cite{Kim2002} and
broadcasting process \cite{Zhou2006}, to enhance network
synchronizability \cite{Motter2005}, to improve traffic capacity
\cite{Yin2006}, and so on. We believe this method can also be
applied to the studies on many other network dynamical processes.

\end{document}